\documentclass[aps,prl,twocolumn,superscriptaddress,floatfix,noeprint]{revtex4-2}
\usepackage[utf8]{inputenc}
\usepackage[colorlinks=true,allcolors=blue]{hyperref}

\usepackage{titlesec}
\usepackage{etoolbox}

\titleformat{\section}[runin]    
  {}{}{}{\hspace{1em}\textsl}    
  [.\textemdash]               
\titlespacing*{\section}{0pt}{0.5ex}{0.3ex}

\usepackage{graphicx}
\usepackage[normalem]{ulem}
\usepackage{xcolor}

\usepackage{amsmath}
\usepackage{amssymb}
\usepackage{bm}
\usepackage{mathtools}
\usepackage{braket}

\DeclarePairedDelimiter{\mean}{\langle}{\rangle}

\newcommand\violetsout{\bgroup\markoverwith{\textcolor{violet}{\rule[0.5ex]{2pt}{0.4pt}}}\ULon}

\begin{document}

\title{Effects of retardation on many-body superradiance in chiral waveguide QED}

\author{Bennet Windt}
\email{bennet.windt@mpq.mpg.de}
\address{Max Planck Institute of Quantum Optics, Hans-Kopfermann-Stra{\ss}e 1, 85748 Garching, Germany}
\address{Munich Center for Quantum Science and Technology, Schellingstra{\ss}e 4, 80799 M{\"u}nchen, Germany}

\author{Miguel Bello}
\email{miguel.bello@mpq.mpg.de}
\address{Max Planck Institute of Quantum Optics, Hans-Kopfermann-Stra{\ss}e 1, 85748 Garching, Germany}
\address{Munich Center for Quantum Science and Technology, Schellingstra{\ss}e 4, 80799 M{\"u}nchen, Germany}

\author{Daniel Malz}
\address{Department of Mathematical Sciences, University of Copenhagen, Universitetsparken 5, 2100 Copenhagen, Denmark}

\author{J. Ignacio Cirac}
\address{Max Planck Institute of Quantum Optics, Hans-Kopfermann-Stra{\ss}e 1, 85748 Garching, Germany}
\address{Munich Center for Quantum Science and Technology, Schellingstra{\ss}e 4, 80799 M{\"u}nchen, Germany}

\begin{abstract}
We study the superradiant decay of a chain of atoms coupled to a chiral waveguide, focusing on the regime of non-negligible photon propagation time. Using an exact master equation description which accounts for delay effects, we obtain evidence to suggest that competition between collective decay and retardation leads to the emergence of an effective maximum number of atoms able to contribute to the superradiant dynamics, resulting in a plateau of the peak emission rate. 
To develop this analysis further, we investigate the inter-atomic correlations to find features consistent with the formation of individual superradiant domains. Moreover, we find that retardation can also result in persistent oscillatory atomic dynamics accompanied by a periodic sequence of emission bursts.
\end{abstract}

\maketitle

A major objective of modern quantum optics is to characterise cooperative phenomena, one of the most striking examples being \emph{superradiance}~\cite{gross_superradiance_1982}, the emission of a short radiation burst from a collection of $N$ emitters, with a peak intensity scaling superlinearly in $N$. Since the first formulation of the theory of superradiance~\cite{dicke_coherence_1954}, there has been significant interest in understanding superradiant dynamics in extended media~\cite{burnham_coherent_1969,arecchi_cooperative_1970,macgillivray_theory_1976,karnyukhin_semiclassical_1982,kaneva_self-organization_1990,kaneva_self-organization_1991}, where the build-up of coherence among atoms may be undermined by their spatial separation. In particular, a key question which remains open is how superradiance is affected in the regime where propagation times of the photons mediating dissipative interactions are non-negligible, rendering the dynamics non-Markovian.

A novel setting for exploring superradiance is that of \emph{waveguide quantum electrodynamics} (WQED). Here, long-range dissipative interactions mediated by photonic modes confined in a waveguide enable the superradiant decay of distant emitters~\cite{chang_colloquium_2018,sheremet_waveguide_2023,cardenas-lopez_many-body_2023}, signatures of which have already been observed in a variety of experimental platforms~\cite{goban_superradiance_2015,sipahigil_integrated_2016,kim_super-radiant_2018,grim_scalable_2019,liedl_collective_2023,liedl_observation_2023,tiranov_collective_2023}. In suitably structured waveguides, the group velocity of the guided modes can be engineered~\cite{baba_slow_2008,lodahl_interfacing_2015} to enter the regime of non-negligible retardation in a controlled manner. Some theoretical works have already explored the non-Markovian effects which arise in this slow-light WQED setting~\cite{gonzalez-ballestero_non-markovian_2013,sinha_non-markovian_2020,pichler_photonic_2016,vodenkova_continuous_2023,ramos_non-markovian_2016,papaefstathiou_efficient_2024}, however these studies were limited to a few excitations, since simulating many-body dynamics beyond the Markovian approximation is generally exceedingly difficult.

As we will demonstrate, this is not true in the case of a \emph{chiral} waveguide, where the photon-mediated interactions are unidirectional~\cite{lodahl_chiral_2017,mitsch_quantum_2014,sollner_deterministic_2015}. Here, the emitter-waveguide setup constitutes a \emph{cascaded quantum system}~\cite{gardiner_driving_1994,carmichael_quantum_1993,Gardiner2000}, which is much more tractable. Recently, a superradiant burst from $N\approx 10^3$ atoms in such a system was experimentally observed~\cite{liedl_observation_2023,bach_emergence_2024}, and chiral WQED therefore presents a unique vantage point for exploring the effects of retardation on superradiance, combining the possibility of rigorous theoretical investigation with a proximity to experiments. 

In this Letter, we show that non-negligible photon propagation times can significantly alter the collective decay dynamics in a chiral WQED setting. Our simulations, based on an exact master equation description, reveal that retardation suppresses the characteristic superradiant scaling of the peak emission rate, leading instead to a plateau in peak emission along the chain of waveguide-coupled atoms. This signals the emergence of a maximum cooperative system size, resulting in local rather than global synchronisation among the atoms. We characterise this further by studying the atomic correlations and supplement our numerical observations by mean-field analyses. Finally, we demonstrate that sufficiently long chains support sustained oscillatory dynamics, resulting in a periodic emission of intensity bursts.

\section*{Model \& exact master equation}
We consider a one-dimensional linear waveguide and $N$ two-level atoms coupled chirally to the right-propagating guided modes, described by the Hamiltonian ($\hbar=1$)~\cite{shen_coherent_2005,chang_cavity_2012,stannigel_driven-dissipative_2012,lodahl_chiral_2017,pichler_quantum_2015}
\begin{equation}
\label{eq:hamiltonian}
\begin{split}
    H(t) &= -iv\int_{-\infty}^\infty dx\,b^\dagger(x)\frac{\partial}{\partial x}b(x) \\
    &\quad\quad\qquad+\sqrt{\Gamma v}\,\Theta(t)\sum_{n=1}^N\left(\sigma_n^\dagger b(x_n)+\mathrm{H.c.}\right),
\end{split}
\end{equation}
where $v > 0$ is the group velocity of the guided modes, $\Gamma$ is the single-atom decay rate into the waveguide, and $x_n$ denotes the position of the $n$th atom. We label the atoms according to their position, with $x_n < x_{n+1}$. We assume here perfect coupling into the waveguide but demonstrate the robustness of our results in the presence of additional decay into unguided modes in the End Matter.

Moving beyond the Markovian assumption of negligible photon propagation times typically requires explicit simulation of the environment. However, in cascaded settings such as the one considered here, the absence of feedback from atoms further downstream allows us to trace out the waveguide exactly. To this end, we introduce a \emph{time-shifted picture}, where the time axis for atomic operators is shifted depending on the atomic position. 
Explicitly, for an arbitrary operator $O_n$ acting on the $n$th emitter, the relation between the Heisenberg picture and the time-shifted picture (denoted with a hat) is $\hat{O}_n(t) = O_n(t - t_n)$, where $t_n = (x_N - x_n)/v$. Alternatively, we can define the atomic density operator in the time-shifted picture, $\hat\rho(t)$, which evolves as (see Supplementary Material (SM)~\cite{supp}),
\begin{equation}
\label{eq:qme}
\begin{split}
    \partial_t\hat\rho(t) &= \sum_n \Theta(t - t_n) \mathcal{L}_n \hat\rho(t) \\
    \mathcal{L}_n \hat\rho(t) &= \frac{\Gamma}{2}[\sigma_n\hat\rho(t),\sigma^\dag_n] + \Gamma\sum_{m > n} [\sigma_n\hat\rho(t), \sigma^\dag_m] + \mathrm{H.c.}
\end{split}
\end{equation}
Although it is well established that delay times can be accounted for through suitable frame transformations~\cite{Gardiner2000}, the physical consequences of this have hardly been explored. In particular, most previous works have focused on steady-state properties or sufficiently long-time dynamics~\cite{pichler_photonic_2016, gardiner_driving_1993, carmichael_quantum_1993, gardiner_driving_1994, cirac_quantum_1997}, for which time delays have no significant impact. However, in the context of superradiance, they play a crucial role.

In the following, we assume the emitters to be equidistantly arranged with spacing $d$, and denote by $\tau=d/v$ the time taken for photons to propagate between neighbouring atoms. We focus on initial states with no photons in the waveguide and the emitters in a pure state $\rho(t=0)=\ket{\psi_0}\!\bra{\psi_0}$, where
\begin{equation}
\label{eq:initial_state}
    \ket{\psi_0}=\left[\cos\left(\frac{\theta_0}{2}\right)\ket{g}+\sin\left(\frac{\theta_0}{2}\right)\ket{e}\right]^{\otimes N}.
\end{equation}
We study the collective decay of these states, after the atom-photon coupling is turned on at time $t=0$, as accounted for by the Heaviside step function $\Theta(t)$ in Eq.~\eqref{eq:hamiltonian}. Specifically, we compute local emission rates $r_n(t)=-\partial_t\mean{\sigma_n^\dagger\sigma_n}_t$, with $\mean{\ldots}_t\equiv\mathrm{tr}(\ldots\rho(t))$, noting that the dynamics of any emitter is independent of the presence of more emitters further downstream, so $r_n(t)$ is insensitive to the total number of atoms $N$.

Even with the emitter-only description of the exact dynamics through Eq.~\eqref{eq:qme}, its simulation still presents a considerable computational challenge. In this work, we tackle this problem using simulations based on the Truncated Wigner Approximation (TWA)~\cite{polkovnikov_phase_2010,schachenmayer_many-body_2015,huber_phase-space_2021,huber_realistic_2022,mink_collective_2023,mink_hybrid_2022} (see SM~\cite{supp} for details). In the large-$N$ limit, Dicke superradiance is known to consist of classical evolution driven by initial quantum fluctuations~\cite{Agarwal1974}, which aligns ideally with the approximations of the TWA~\cite{degiorgio_approximate_1971,degiorgio_statistical_1970,bonifacio_cooperative_1975,bonifacio_cooperative_1975-1,haake_fluctuations_1979,haake_quantum_1972,glauber_superradiant_1976}. Since chiral interactions and time delays break the permutation symmetry of the Dicke problem, we further establish the accuracy of our TWA simulations by benchmarking against quantum trajectory simulations with Matrix Product States (MPS), finding excellent agreement (see SM~\cite{supp}).

\begin{figure}[t!]
    \centering
    \includegraphics[width=\linewidth]{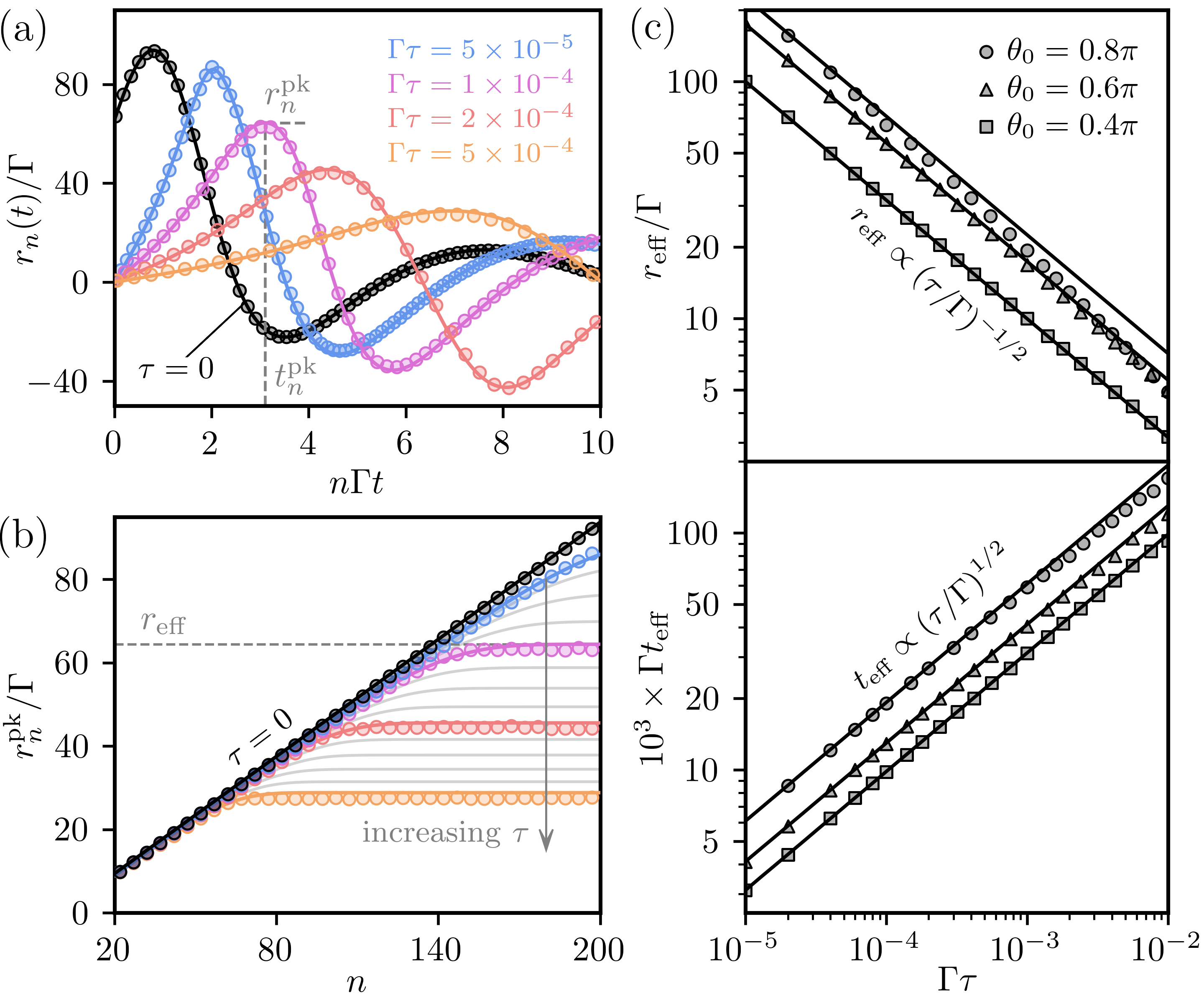}
    \caption{Effect of retardation for a partially-inverted state with $\theta_0=0.7\pi$, computed using TWA (markers) and MFT (solid lines). (a) Local emission rate $r_n(t)$ at $n=200$, evaluated for various time delays $\tau$. (b) Scaling of $r^\mathrm{pk}_n$ with $n$ for various $\tau$, with the values of $\tau$ shown in panel (a) indicated by the corresponding colours. (c) Comparison of the peak emission rate $r_\mathrm{eff}$ and time $t_\mathrm{eff}$ with the analytical predictions~\eqref{eq:plateau_scalings}.}
    \label{fig:partially-inverted}
\end{figure}

\begin{figure*}[t!]
    \centering
    \includegraphics[width=\linewidth]{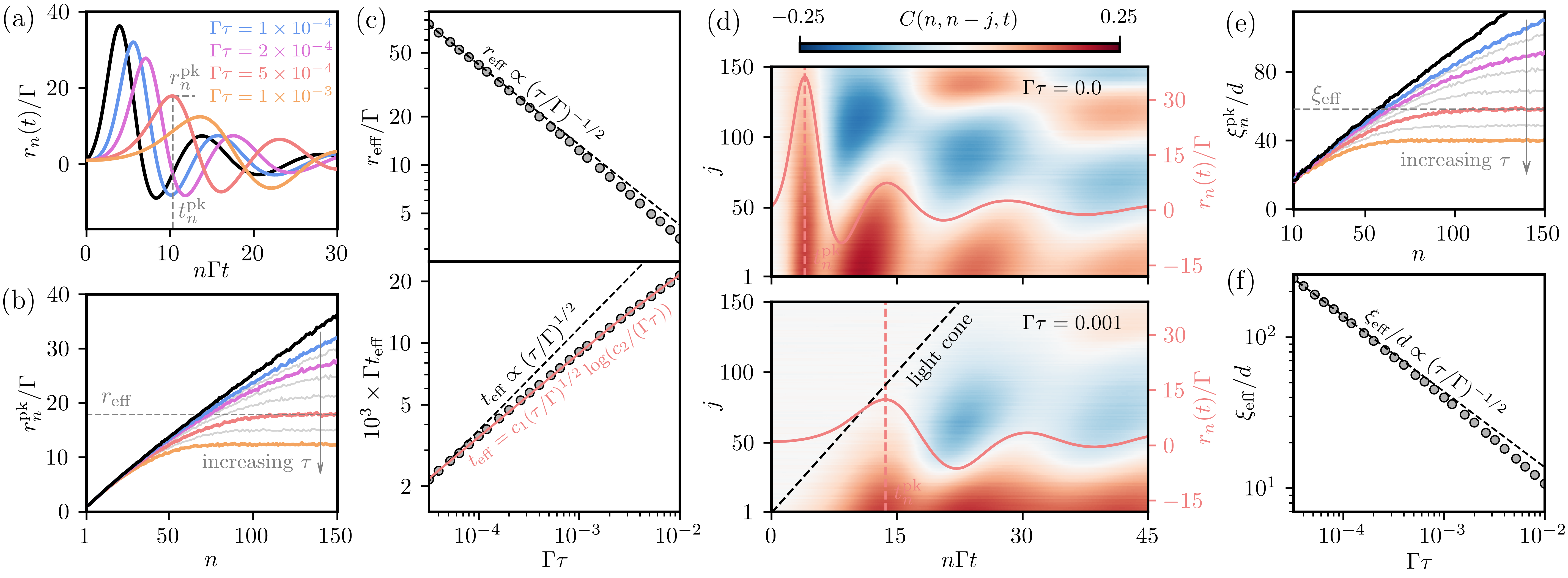}
    \caption{(a)-(c) Same as Figs.~\ref{fig:partially-inverted}a-c for a fully-inverted initial state, with $n=150$ in (a). (d) Time-evolution of the correlator $C(n,n-j,t)$ at $n=150$, for $\tau=0$ (top) and $\Gamma\tau=10^{-3}$ (bottom). In both panels, we overlay $r_n(t)$ for qualitative reference and also indicate the peak emission time $t^\mathrm{pk}_n$. In the lower panel, we also indicate the light cone. (e) Scaling of $\xi^\mathrm{pk}_n$ with $n$ for increasing $\tau$, with the values of $\tau$ shown in panel (a) indicated by the corresponding colours. (f) Scaling of the asymptotic correlation length $\xi_\mathrm{eff}$ with $\tau$, with the small-$\tau$ power law scaling indicated by the dashed line.}
    \label{fig:fully-inverted}
\end{figure*}

\section*{Partially-inverted initial states}
In Fig.~\ref{fig:partially-inverted}a, we show emission dynamics for a partially-inverted initial state ($\theta_0<\pi$ in Eq.~\eqref{eq:initial_state}) for different time delays $\tau$. We observe the characteristic signature of a superradiant burst, namely a peak value $r_n^\mathrm{pk}$ of the site-$n$ emission rate $r_n(t)$ exceeding the single-atom maximum value, $r^\mathrm{pk}_n>\Gamma$, at a delayed peak emission time $t=t^\mathrm{pk}_n>0$. With respect to the retardation-free case, the effect of the retardation is to delay the peak emission further, resulting in a superradiant burst with a lower amplitude $r^\mathrm{pk}_n$ at a later time $t^\mathrm{pk}_n$. This reflects the fact that a non-negligible retardation time allows for atoms downstream to decay independently before the light emitted from previous atoms impinges on them, reducing the buildup of the superradiant pulse along the waveguide. Note also that $r_n(t=0)>\Gamma$ for $\tau=0$, since a state~\eqref{eq:initial_state} with $\theta_0<\pi$ contains (classical) correlations between atoms, however for $\tau>0$, their effect can only manifest progressively. 

In Fig.~\ref{fig:partially-inverted}b, we show the scaling of the peak emission rate $r^\mathrm{pk}_n$ with $n$. In the retardation-free case, we observe the Dicke scaling $r^\mathrm{pk}_n\sim n\Gamma$~\cite{dicke_coherence_1954,gross_superradiance_1982}, while for $\tau>0$, the value of $r^\mathrm{pk}_n$ instead plateaus to a maximum effective emission rate $r_\mathrm{eff}$. A similar effect can be seen for the peak emission time $t^\mathrm{pk}_n$, which plateaus to some $t_\mathrm{eff}$. 
This suggests that the competition between retardation and cooperative emission leads to an effective maximum number of atoms $N_\mathrm{eff}$ which can contribute to the superradiant decay. This maximum cooperative system size can be estimated by a self-consistency argument: the time taken for photons emitted by the first atom to reach the $n$-th atom along the waveguide is $n\tau$, while this atom decays superradiantly on a timescale $\sim 1/(n\Gamma)$. Equating these times, the effective number of atoms upstream from the $n$th atom which can enhance its emission should scale as $N_\mathrm{eff}\sim (\Gamma\tau)^{-1/2}$, such that the Dicke relations $r_n^\mathrm{pk}\sim n\Gamma$ and $t_n^\mathrm{pk}\sim 1/(n\Gamma)$ imply
\begin{equation}
\label{eq:plateau_scalings}
    r_\mathrm{eff}\sim (\tau/\Gamma)^{-1/2}\quad\text{and}\quad t_\mathrm{eff}\sim(\tau/\Gamma)^{1/2}\,.
\end{equation}
While these scalings were already proposed in Ref.~\cite{arecchi_cooperative_1970}, we are now in a position to confirm them numerically. In Fig.~\ref{fig:partially-inverted}c, we show the scalings of $r_\mathrm{eff}$ and $t_\mathrm{eff}$ with $\tau$, and find that, indeed, at sufficiently small $\Gamma\tau$, the scalings~\eqref{eq:plateau_scalings} are observed. 
At larger delays, however, we find that for sufficiently large initial inversion angles $\theta_0$, the scalings diverge from the predicted scalings. This can be understood by noting that the scalings $r_n^\mathrm{pk}\sim n\Gamma$ and $t_n^\mathrm{pk}\sim1/(n\Gamma)$ hold for $n\gg1$, however at large $\Gamma\tau$ the effective cooperative system size $N_\mathrm{eff}$ becomes small enough for these scalings to break down. This effect is more pronounced for larger $\theta_0$, where the asymptotic scalings are approached more slowly.

\section*{Mean-field analysis}
For partially-inverted initial states, the emission dynamics are also captured by a mean field theory (MFT) description in the limit of small delay times ($\Gamma\tau\ll1$), where the atomic array can be treated as a continuous medium~\cite{burnham_coherent_1969,karnyukhin_semiclassical_1982,arecchi_cooperative_1970,haake_fluctuations_1979,kaneva_self-organization_1990,kaneva_self-organization_1991,gross_superradiance_1982,bonifacio_steady-state_1975,hopf_quantum_1976}. We express $\mean{\sigma_n}_t=\frac{1}{2}\sin\theta(n,\Gamma t)$ and $\mean{\sigma_n^z}_t=-\frac{1}{2}\cos\theta(n,\Gamma t)$, in terms of a continuous bivariate function $\theta(x,t)$ with dimensionless arguments obeying (see SM~\cite{supp} and Refs.~\cite{mahmoodian_dynamics_2020,calajo_emergence_2022})
\begin{equation}
\label{eq:mft}
    \left[\left(\Gamma\tau\right)\frac{\partial^2}{\partial t^2}+\frac{\partial^2}{\partial t\partial x}\right]\theta(x,t)+\sin\theta(x,t)=0\,,
\end{equation}
with boundary conditions $\theta(x,0)=\theta(0,t)=\theta_0$. A numerical solution using the method of finite differences~\cite{LeDret2016} shows excellent agreement of the emission dynamics with the TWA numerics for sufficiently small delays (see Figs.~\ref{fig:partially-inverted}a,b).

Moreover, Eq.~\eqref{eq:mft} itself serves as a starting point for some insightful scaling arguments: For $\tau=0$, Eq.~\eqref{eq:mft} becomes the \emph{Sine-Gordon equation}~\cite{Drazin1989} $\partial_t\partial_x\theta(x,t)+\sin\theta(x,t)=0$, which is symmetric under $x\leftrightarrow t$. With our boundary conditions, it can be solved as $\theta(x,t)=u(xt)$ with $z\,u''(z)+u'(z)+\sin u(z)=0$, subject to the initial conditions $u(z=0)=\theta_0$ and $u'(z=0)=-\sin\theta_0$~\cite{arecchi_cooperative_1970}. Note that this change of variables implies $\theta_n(t)=\theta_m(nt/m)$ and hence $r_n(t)=(n/m)\,r_m(nt/m)$, which is a statement of the scalings for the retardation-free case assumed above. More generally, we can see from Eq.~\eqref{eq:mft} that the solution $\theta(x, t; \tau)$ for a given delay $\tau$ is related to the solution for a delay $\alpha\tau$ ($\alpha>0$) as $\theta(x, t; \alpha\tau) = \theta(\sqrt{\alpha}x, t/\sqrt{\alpha};\tau)$. Defining the associated emission rate $r(x,t;\tau)=-\frac{1}{2}\sin\theta(x,t;\tau)\partial_t\theta(x,t;\tau)$, we can then see that $r(x,t;\alpha\tau)=r(\sqrt{\alpha}x, t/\sqrt{\alpha};\tau)/\sqrt{\alpha}$. From these relations, it follows directly that the asymptotic scalings in Eq.~\eqref{eq:plateau_scalings} are exact at the mean-field level (see Fig.~\ref{fig:partially-inverted}c) and the departure from these scalings at larger $\Gamma\tau$ is consistent with the breakdown of the continuum approximation in this regime.

\section*{Fully-inverted initial state}
Beyond capturing departures from mean-field predictions, our numerics also allow us to study the fully-inverted initial state $\ket{\psi_0}=\ket{e}^{\otimes N}$ conventionally considered in Dicke superradiance, which is adynamical under Eq.~\eqref{eq:mft}, since the decay of this state is initialised by vacuum fluctuations not accounted for by MFT~\cite{Agarwal1974}. In Figs.~\ref{fig:fully-inverted}a,b we see that the plateau effect described above for partially-inverted states manifests analogously in this case. However, the scaling of $t_\mathrm{eff}$ displays a significant divergence from the predicted scaling. This can be understood by noting that the correct $n$-dependence of $t^\mathrm{pk}_n$ in the retardation-free case is actually given
by $t^\mathrm{pk}_n=c_1\log(c_2n)/(n\Gamma)$, with some constants $c_1,c_2$~\cite{malz_large-n_2022}. Indeed, we find that the scaling $t_\mathrm{eff}=c_3(\tau/\Gamma)^{1/2}\log(c_4/(\Gamma\tau))$ implied by this is a much better fit (see Fig.~\ref{fig:fully-inverted}c).

We now consider the implications of the emergent maximum cooperative system size $N_\mathrm{eff}$ for a chain of $N\gg N_\mathrm{eff}$ atoms: intuitively, such a chain can be expected to ``split'' into individual regions of size $\sim N_\mathrm{eff}$, each decaying superradiantly. As a starting point for characterising this domain structure, we introduce the multi-time correlator
\begin{equation}
\label{eq:correlator}
    C(n,m,t)=\mean{\sigma_n^\dagger(t)\sigma_m(t-(n-m)\tau)}_0\,,
\end{equation}
which is real-valued for the fully-inverted initial state. In Fig.~\ref{fig:fully-inverted}d, we show the time-evolution of $C(n,n-j,t)$. In the retardation-free case, we observe a buildup of correlations around $t\approx t^\mathrm{pk}_n$, reflecting the emergent coherence across the atoms known to accompany superradiant decay~\cite{gross_superradiance_1982,dicke_coherence_1954}. Contrary to the Dicke case, however, these correlations are not uniform, but decrease with distance $j$ from the $n$th site.

The case of $\tau>0$ displays two important differences with respect to the retardation-free case: Firstly, Fig.~\ref{fig:fully-inverted}d shows the emergence of a ``light cone'' structure, since $C(n,n-j,t)=0$ for $t<j\tau$. Secondly, the correlations at peak emission decay with $j$ over shorter length scales along the atomic array. To make this second observation more quantitative, we note that  for $t\leq t^\mathrm{pk}_n$, the correlations are well-captured by a compressed exponential fit $C(n,n-j,t)=C_0e^{-(jd/\xi_n(t))^\alpha}$ (see SM~\cite{supp}), where the fit parameter $\xi_n(t)$ can be interpreted as a (time-dependent) correlation length. In the retardation-free case, we find that its value at $t=t^\mathrm{pk}_n$, which we denote by $\xi^\mathrm{pk}_n=\xi_n(t=t_n^\mathrm{pk})$, scales as $\xi^\mathrm{pk}_n\sim nd$. For $\tau>0$, we can see that $\xi^\mathrm{pk}_n$ instead approaches a fixed asymptotic value $\xi_\mathrm{eff}$ just like $r^\mathrm{pk}_n$ and $t^\mathrm{pk}_n$ (see Fig.~\ref{fig:fully-inverted}e). This asymptotic $n$-independent correlation length provides a direct signature of the formation of superradiant domains: it implies that in a large system, any atom sufficiently far along the chain can synchronise only across a finite range $\sim\xi_\mathrm{eff}$, leading to \emph{local} rather than global collective decay. Additionally, we find that $\xi_\mathrm{eff}$ displays the scaling $\xi_\mathrm{eff}/d\sim(\Gamma\tau)^{-1/2}$ at sufficiently small $\tau$ (see Fig.~\ref{fig:fully-inverted}f), in accordance with our interpretation of a maximum cooperative system size.

\section*{Multi-peak structure}
\begin{figure}[t!]
    \centering
    \includegraphics[width=\linewidth]{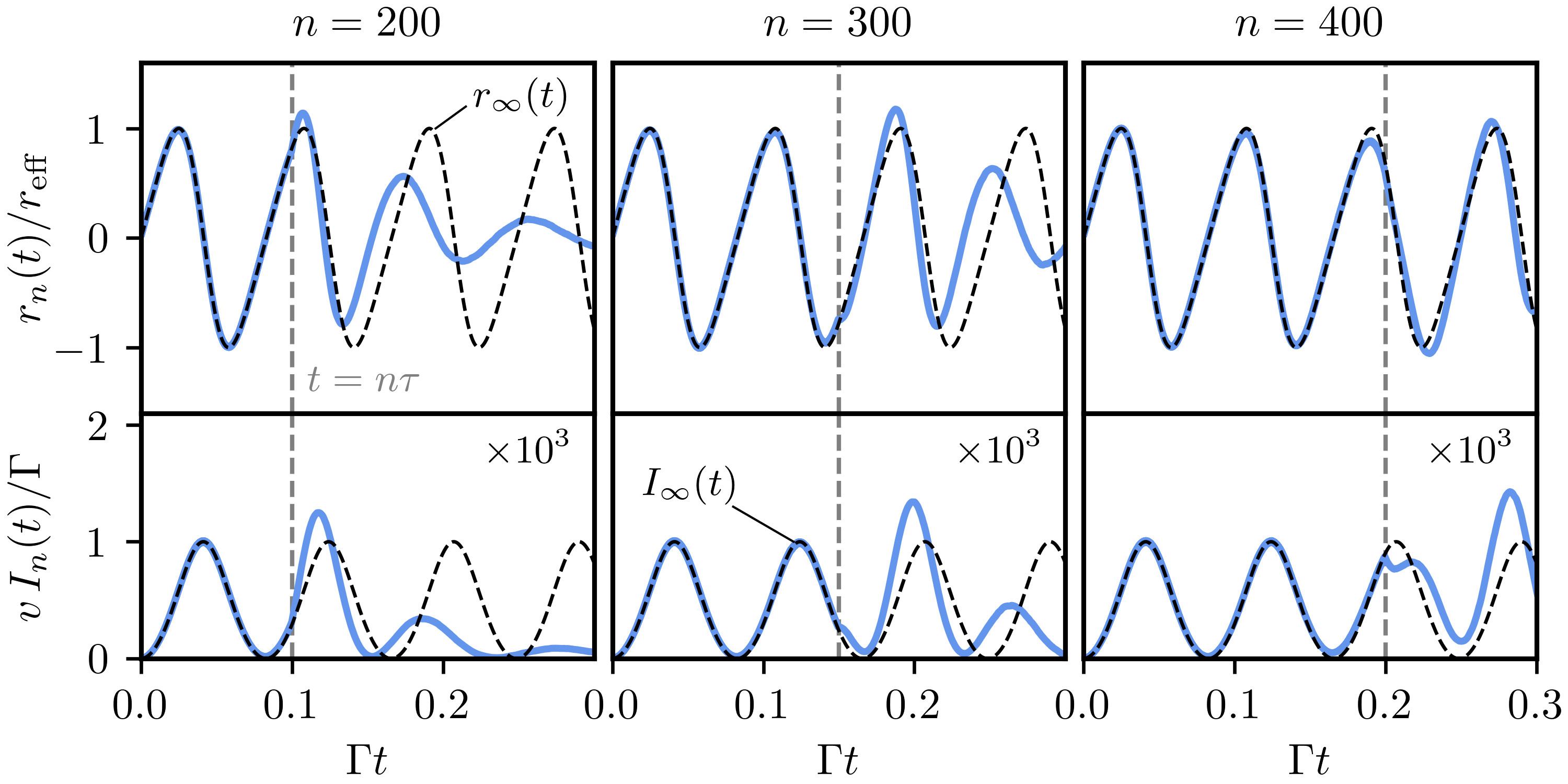}
    \caption{Dynamics of the site-$n$ emission rate $r_n(t)$ (top row) and local field intensity $I_n(t)$ (bottom row) for $\theta_0=\pi/2$ and $\Gamma\tau=5\times10^{-4}$, evaluated at various $n$ using TWA (solid blue lines). We also indicate the asymptotic mean-field predictions (dashed black lines) and the time $t=n\tau$ (dashed gray lines).}
    \label{fig:multi-peaks}
\end{figure}

Finally, while in the preceding discussion we have focused entirely on the properties of the first maximum of $r_n(t)$, retardation also leads to intriguing novel features in the emission dynamics at later times. For $\tau>0$, our preceding discussion has shown that the evolution of all atoms sufficiently far down the chain will be determined by an identical environment of a fixed number of atoms further upstream. At the level of the MFT, this can be formalised as $\theta(x\to\infty,t)=\theta_\infty(t)$, where $\theta_\infty(t)$ denotes the solution to the equation obtained by dropping the spatial derivative in Eq.~\eqref{eq:mft}, i.e. $(\Gamma\tau)\ddot{\theta}_\infty(t)+\sin\theta_\infty(t)=0$. The solution to this equation can be obtained analytically (see SM~\cite{supp}) and displays a \emph{periodic} time-dependence: $\theta_\infty(t)$ oscillates between $\pm\theta_0$, while the asymptotic (uniform) local emission rate $r(x\to\infty,t)=-\frac{1}{2}\sin\theta_\infty(t)\dot{\theta}_\infty(t)\equiv r_\infty(t)$ oscillates between $\pm r_\mathrm{eff}$. In Fig.~\ref{fig:multi-peaks}, we compare this mean-field prediction against our numerics by examining the site-$n$ emission rate $r_n(t)$ for $n\gg 1$. We observe that indeed $r_n(t)\approx r_\infty(t)$ until a time $t\sim n\tau$, at which point the finite-site effects propagating from the beginning of the emitter chain reach the $n$th site. A similar $n$-independent periodic behaviour can be seen for the field intensity in the waveguide immediately after the $n$th site,  $I_n(t)=\mean{b^\dagger(x_n^+)b(x_n^+)}_t$ (with $x^+\equiv x+0^+$), which can be expressed in terms of local emission rates as
\begin{equation}
\label{eq:intensity}
    I_n(t)=\frac{1}{v}\sum_{m\leq n}\Theta(t-(n-m)\tau)\,r_m(t-(n-m)\tau)\,.
\end{equation}
Specifically, the oscillations in $r_n(t)$ are accompanied by a sequence of emission bursts in $I_n(t)\approx I_\infty(t)$ (see Fig.~\ref{fig:multi-peaks}).

The prediction of infinitely sustained periodic dynamics in the thermodynamic limit raises intriguing questions. Firstly, the precise physical mechanisms underlying the phenomenon remain to be understood. Secondly, the fate of the oscillations beyond MFT is unclear, since the errors due to neglected correlations at the mean-field level accumulate over time, and in a many-body system like the one considered here, thermalisation effects could be expected to wash out coherent dynamics at late times. We leave a detailed investigation of these questions to future work, and emphasise here only that the dynamics in Fig.~\ref{fig:multi-peaks} represent a departure from the well-studied retardation-free superradiant dynamics.

\section*{Conclusion}
In this Letter, we have shown that non-negligible photon propagation times can lead to striking qualitative departures from the standard theory of superradiance, by inhibiting the global synchronisation of the atoms typically associated with superradiant decay. An effective maximum cooperative system size emerges, which manifests in a suppression of the characteristic superlinear scaling of the peak emission rate with system size, and in the formation of locally synchronised superradiating domains. Building on a mean-field description, we have also identified regimes in which retardation induces sustained periodic atomic dynamics which manifest in a regular sequence of emission bursts.

The contribution of our work is two-fold: Firstly, we have provided a partial answer to the long-standing problem of superradiance in extended atomic media, relying on sophisticated numerical simulations beyond qualitative conjectures. In doing so, we have also identified the setting of slow-light chiral WQED as a potential experimental platform for exploring this regime in a controlled manner. Secondly, our results provide a first glimpse into the varied phenomena that could be observed in collective decay dynamics beyond the Markov approximation. Suitably structured photonic environments can give rise to a host of non-Markovian phenomena, which have only been explored in the single-excitation regime~\cite{gonzalez-tudela_exotic_2018,gonzalez-tudela_quantum_2017,redondo-yuste_quantum_2021,di_benedetto_dipole-dipole_2024,lambropoulos_fundamental_2000,kim_super-_2023}, and our work is a timely addition to the nascent research effort to explore the many-body dynamics of such systems.

\section*{Acknowledgements}
We are very grateful to C. Mink for generous advice on the implementation of the TWA numerics. B.W., M.B., and J.I.C. acknowledge funding from the Munich Center for Quantum Science and Technology (MCQST), funded by the Deutsche Forschungsgemeinschaft (DFG) under Germany’s Excellence Strategy (EXC2111-390814868).  D. M. acknowledges support from the Novo Nordisk Foundation under grant numbers NNF22OC0071934 and NNF20OC0059939.

\section*{Note added}
During the completion of this manuscript, a related work applying the TWA to superradiant decay in chiral WQED appeared in Ref.~\cite{tebbenjohanns_predicting_2024}, which however focuses exclusively on the retardation-free case.

\begin{center}
    \medskip\textbf{END MATTER}\medskip
\end{center}

In realistic settings, the waveguide-coupled atoms are typically also subject to independent decay into unguided free-space modes, which competes with the correlated emission into the waveguide. We account for this effect by supplementing the unitary dynamics under the Hamiltonian~\eqref{eq:hamiltonian} with a local dissipation term $\gamma\sum_n\mathcal{D}[\sigma_n]$ with $\mathcal{D}[x]~\boldsymbol{\cdot}=x\boldsymbol{\cdot}x^\dagger-\{x^\dagger x,\boldsymbol{\cdot}\}/2$. At the level of the exact master equation~\eqref{eq:qme}, this amounts to the modification $\mathcal{L}_n \hat\rho(t)\to \mathcal{L}_n \hat\rho(t)+\gamma\mathcal{D}[\sigma_n]\hat\rho(t)$.
Here, we discuss the effect of this additional loss mechanism on the retardation-induced peak emission plateau. 

\begin{figure}[t!]
    \centering
    \includegraphics[width=\linewidth]{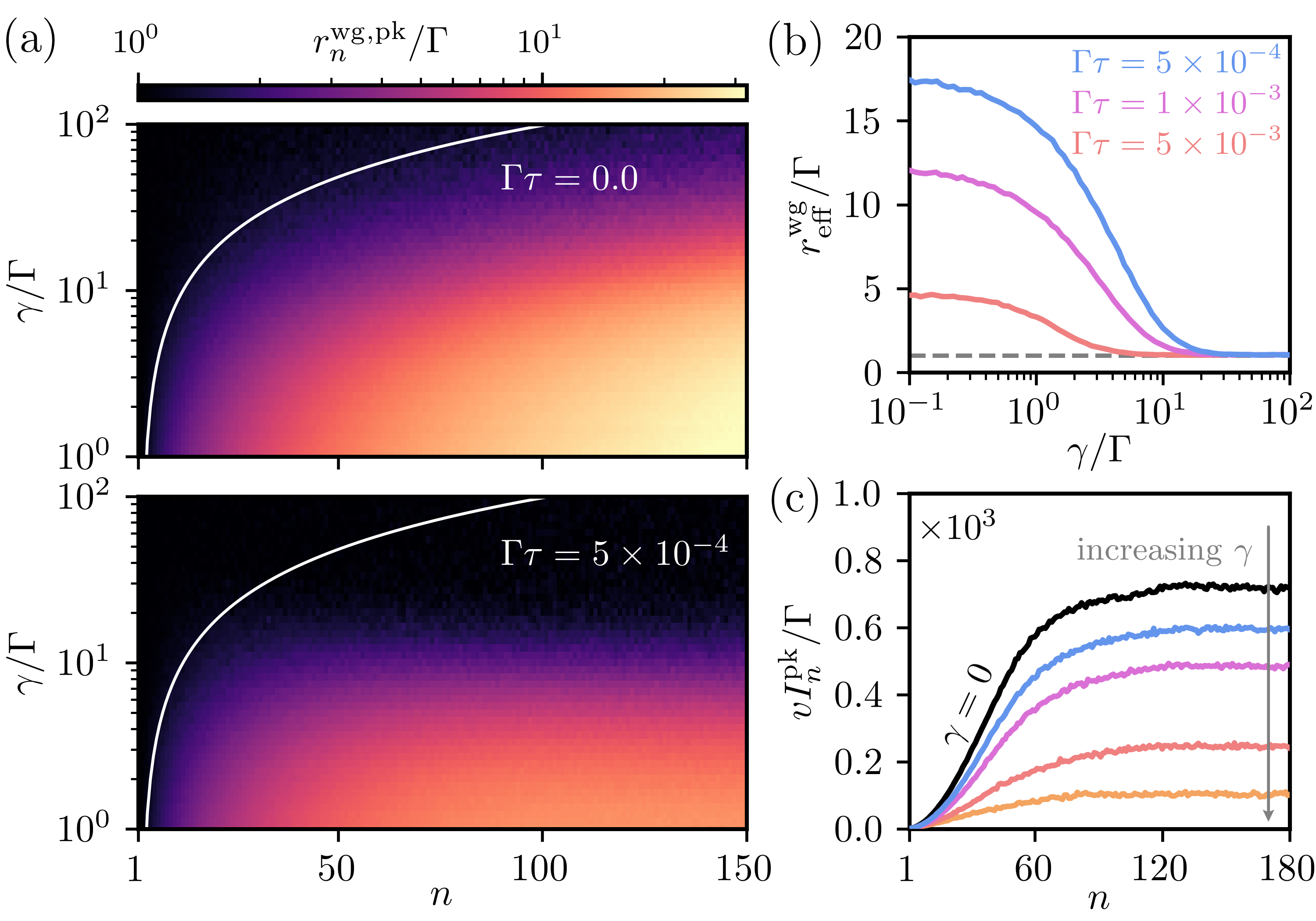}
    \caption{(a) Peak emission rate into the waveguide for $\tau=0$ (top) and $\Gamma\tau=5\times10^{-4}$ (bottom). The retardation-free superradiance criterion from Ref.~\cite{cardenas-lopez_many-body_2023} is indicated in white. (b) Plateau emission rate $r_\mathrm{eff}^\mathrm{wg}$ as a function of free-space decay rate $\gamma$ for various time delays. (c) Local field intensity along the waveguide for $\Gamma\tau=10^{-3}$, both without free-space decay and with $\gamma/\Gamma\in\{1,2,5,10\}$, as indicated by the arrow.}
    \label{fig:free-space}
\end{figure}

Specifically, we consider local emission rates \emph{into the waveguide}, which we define as $r_n^\mathrm{wg}(t)=r_n(t)-\gamma\mean{\sigma_n^\dagger\sigma_n}_t$. In the absence of retardation, superradiant decay of the $n$-th atom (i.e. a peak local emission rate $r_n^\mathrm{wg,pk}>\Gamma$ into the waveguide) can only occur for $n\geq\gamma/\Gamma+2$~\cite{cardenas-lopez_many-body_2023}, as we confirm in Fig.~\ref{fig:free-space}a. Beyond this threshhold value, the asymptotic Dicke scaling $r_n^\mathrm{wg,pk}\sim n\Gamma$ is recovered for sufficiently large $n$.

When $\tau>0$, we observe a more intricate interplay between collective and independent emission, owing to the additional loss of excitations into unguided modes during the propagation time of the photons in the waveguide. As shown in Fig.~\ref{fig:free-space}a, at sufficiently small time delays the burst criterion from the retardation-free regime still holds and at sufficiently large $n$ the plateau effect can be observed. However, the plateau emission rate $r^\mathrm{wg}_\mathrm{eff}$ varies with $\gamma$ and, most notably, there exists a critical free-space decay rate $\gamma_\mathrm{max}$ that depends on the retardation time $\tau$, such that for larger $\gamma>\gamma_\mathrm{max}$, the burst vanishes entirely for all values of $n$ and $r_n^\mathrm{wg,pk}=\Gamma$ (see Fig.~\ref{fig:free-space}b).

It is worth noting that the emission rates $r_n^\mathrm{wg}(t)$ into the waveguide are not typically directly measurable in an experimental context. The local field intensity $I_n(t)$ in the waveguide (defined as in Eq.~\eqref{eq:intensity} of the main text, but with $r_n\to r_n^\mathrm{wg}$), represents a more experimentally accessible quantity. In Fig.~\ref{fig:free-space}c, we show that the qualitative signatures of the plateau effect in the peak intensity $I_n^\mathrm{pk}$ are robust even in the presence of comparatively strong free-space decay.

\bibliography{bibliography.bib}

\end{document}


\title{Supplementary Material: \\ Effects of retardation on many-body superradiance in chiral waveguide QED}

\author{Bennet Windt}
\address{Max Planck Institute of Quantum Optics, Hans-Kopfermann-Stra{\ss}e 1, 85748 Garching, Germany}
\address{Munich Center for Quantum Science and Technology, Schellingstra{\ss}e 4, 80799 M{\"u}nchen, Germany}

\author{Miguel Bello}
\address{Max Planck Institute of Quantum Optics, Hans-Kopfermann-Stra{\ss}e 1, 85748 Garching, Germany}
\address{Munich Center for Quantum Science and Technology, Schellingstra{\ss}e 4, 80799 M{\"u}nchen, Germany}

\author{Daniel Malz}
\address{Department of Mathematical Sciences, University of Copenhagen, Universitetsparken 5, 2100 Copenhagen, Denmark}

\author{J. Ignacio Cirac}
\address{Max Planck Institute of Quantum Optics, Hans-Kopfermann-Stra{\ss}e 1, 85748 Garching, Germany}
\address{Munich Center for Quantum Science and Technology, Schellingstra{\ss}e 4, 80799 M{\"u}nchen, Germany}

\maketitle

\setcounter{tocdepth}{1}
\tableofcontents

\section{Exact master equation in time-shifted picture}

The goal of this section is to derive the master equation in the main text. The starting point for this derivation is the Heisenberg equation of motion under the atom-waveguide Hamiltonian for an arbitrary emitter operator $O(t)$,
\begin{equation}
\label{eq:dotO}
    \dot{O}(t)=i\sqrt{\Gamma v}\,\Theta(t)\sum_n\left([\sigma_n^\dagger(t),O(t)]b(x_n,t)+b^\dagger(x_n,t)[\sigma_n(t),O(t)]\right)\,.
\end{equation}
Note that the Hamiltonian can also be expressed in terms of waveguide modes $b(k) = \int_{-\infty}^\infty dx\,e^{-ikx}b(x)/\sqrt{2\pi}$ with well-defined momenta $k$ as $H(t) = H_\mathrm{wg} + \Theta(t)\sum_{n = 1}^N H_n$, where
\begin{equation}
    H_\mathrm{wg} = v \int_{-\infty}^\infty dk \, k \, b^\dagger(k) b(k)\quad\text{and}\quad 
    H_n = \sqrt{\frac{\Gamma v}{2\pi}} \int_{-\infty}^\infty dk \left(e^{ikx_n} \sigma^\dagger_n b(k) + \mathrm{\mathrm{H.c.}}\right) \,. \label{eq:hamiltonianmomentum}
\end{equation}
From this expression, we can derive the Heisenberg equations of motion for the waveguide modes,
\begin{equation}
    \partial_t b(k, t) = -i v k\,b(k, t) - i\sqrt{\frac{\Gamma v}{2\pi}} \, \Theta(t) \sum_n e^{-ik x_n} \sigma_n(t) \,,
\end{equation}
which can be integrated with respect to time to obtain
\begin{equation}
    b(k, t) = e^{-ivkt}\,b(k, 0) - i\sqrt{\frac{\Gamma v}{2\pi}} \sum_n e^{-ikx_n} \int_0^t ds\, e^{-ivk(t - s)}\sigma_n(t) \,.
\end{equation}
The time-evolved real-space modes can be obtained by means of a Fourier transform, $b(x, t) = \int_{-\infty}^\infty dk\,e^{ikx} b(k, t)/\sqrt{2\pi}$.
Evaluated at the emitter positions $x=x_n$, they are given by~\cite{chang_single-photon_2007,chang_cavity_2012,caneva_quantum_2015}
\begin{equation}
    b(x_n, t) = b(x_n - v t, 0) - \frac{i}{2}\sqrt{\frac{\Gamma}{v}}\, \Theta(t) \sigma_n(t) - i\sqrt{\frac{\Gamma}{v}} \sum_{m < n} \Theta(t - t_m + t_n) \sigma_m(t - t_m + t_n) \,.
    \label{eq:fieldevolution}
\end{equation}
The first term corresponds to the free evolution of the waveguide modes, i.e., in the absence of emitters. 
For a nonzero coupling $\Gamma > 0$, the field at position $x_n$ starts to depend on the state of the atoms further upstream after the corresponding delay times, $t_{mn} = t_m - t_n = (x_n - x_m)/v$, for $1\leq m \leq n$. Substituting Eq.~\eqref{eq:fieldevolution} into Eq.~\eqref{eq:dotO}, we realize that the evolution of an operator on the $n$th emitter, $O_n(t)$, depends on operators of other emitters further upstream, evaluated at earlier times,
\begin{equation} 
    \partial_t O_n(t) = i\sqrt{\Gamma v}\, \Theta(t) \left([\sigma^\dagger_n(t), O_n(t)]\left\{b(x_n - vt, 0) - \frac{i}{2}\sqrt{\frac{\Gamma}{v}} \sigma_n(t) - i\sqrt{\frac{\Gamma}{v}} \sum_{m < n} \Theta(t - t_{mn})\sigma_m(t - t_{mn})\right\} + \cdots \right) \,. \label{eq:dotOn}
\end{equation} 
The different time-dependencies can be neatly accommodated by defining time-shifted local operators, $\hat O_n(t) = O_n(t - t_n)$. 
These operators in the time-shifted picture obey
\begin{equation} 
    \partial_t \hat O_n(t) = i\sqrt{\Gamma v}\, \Theta(t - t_n) \left([\hat\sigma^\dagger_n(t), \hat O_n(t)]\left\{b(x_N - vt, 0) - \frac{i}{2}\sqrt{\frac{\Gamma}{v}} \hat \sigma_n(t) - i\sqrt{\frac{\Gamma}{v}} \sum_{m < n} \Theta(t - t_m)\hat \sigma_m(t)\right\} + \cdots \right) \,, \label{eq:dotOntilde}
\end{equation}
where now all atomic operators are evaluated at the same time. 
One can think of the definition of time-shifted operators as a frame transformation to the \emph{time-shifted picture}, where operators evolve as
\begin{equation}
    \hat{O}(t) = V^\dagger(t) O V(t) \,, \quad V(t) = \mathcal{T}_\rightarrow\left\{\exp\left(i\int_0^t H'(s) ds\right)\right\}\,.
\end{equation}
Here, $O$ is an arbitrary operator in the Schrödinger picture, and $\hat{O}(t)$ is the corresponding operator in the time-shifted picture. 
Moreover, $H'(t)$ is a time-dependent Hamiltonian in which emitters are coupled to the waveguide consecutively, starting with the last emitter and ending with the first emitter, 
\begin{equation}
    H'(t) = H_\mathrm{wg} + \sum_n \Theta(t - t_n)H_n \,,
\end{equation}
and $\mathcal{T}_\rightarrow\left\{\dots\right\}$ denotes the \emph{anti-time-ordered} product of the operators inside the curled brackets, such that
\begin{equation}
    V(t) = \begin{cases} 
        U_N(t) \,, & 0 < t \leq t_{N - 1} \\
        U_N(t_{N - 1})U_{N - 1}(t - t_{N - 1}) \,, & t_{N - 1} < t \leq t_{N - 2} \\
        \hspace*{4.5cm}\vdots \\
        U_N(t_{N - 1}) U_{N - 1}(t_{N-2} - t_{N-1}) \cdots U_1(t - t_1) \,, & t > t_1
    \end{cases} \,,
\end{equation}
where $U_m(t) = \exp(-i t(H_\mathrm{wg} + \sum_{j\geq m}H_j))$.
From Eq.~\eqref{eq:dotOn}, we can understand how this transformation produces the desired result when applied to local atomic operators. 
First of all, any $O_n$ commutes with all $U_m(t)$ with $m > n$, so $\hat{O}_n(t) = O_n$ for $t \leq t_n$.
Second, since any $O_n(t)$ starts to depend on $m$th-emitter operators ($m \leq n$) only for $t > t_{mn}$,
\begin{equation}
    U^\dagger_m(t) O_n U_m(t) = U^\dagger_{m-1}(t) O_n U_{m-1}(t) \,,\ \text{for $0 < t \leq t_{m-1} - t_n$ and $m \leq n$} \,.
\end{equation}
All the different Eqs.~\eqref{eq:dotOntilde}, for $n=1, \dots, N$, can be combined into a single one, which reads
\begin{equation}
    \partial_t\hat{O}_n(t) = i\sqrt{\Gamma v}\sum_m \Theta(t - t_m) \left([\hat\sigma^\dag_m(t), \hat{O}_n(t)]b(x_m, t - t_m) + b^\dag(x_m, t - t_m)[\hat\sigma_m(t), \hat{O}_n(t)]\right) \,,
    \label{eq:dotOtilde}
\end{equation}
noting that local, time-shifted operators of different atoms evaluated at the same time commute, $[\hat{O}_m(t), \hat{O}_n(t)] = 0$ for all $m, n$.
In fact, Eq.~\eqref{eq:dotOtilde} holds for arbitrary (possibly non-local) atomic operators in the time-shifted picture, $\hat{O}(t) = \hat{O}_1(t)\hat{O}_2(t)\cdots\hat{O}_N(t)$, noting that for $t > t_m$,
\begin{align*}
    [\hat{O}_n(t), b(x_m, t - t_m)] & = [O_n(t - t_n), b(x_m, t - t_m)] \\
    & = 
    \begin{cases}
        U^\dag_1(t - t_n)[O_n(0), b(x_m, t_{nm})]U_1(t - t_n) \,, & t_n > t_m \\
        U^\dag_1(t - t_m)[O_n(t_{mn}), b(x_m, 0)]U_1(t - t_n) \,, & t_m > t_n \\
    \end{cases} \\
    & = 0 \,,
\end{align*}
where the last equality follows from the expression of the waveguide field, Eq.~\eqref{eq:fieldevolution}, and the equations of motion of local atomic operators, Eqs.~\eqref{eq:dotOn}. 
For $t_n > t_m$, $b(x_m, t)$ only starts to depend on any $O_n(0)$ for $t > t_{nm}$; and, similarly, for $t_m > t_n$, $O_n(t)$ starts to depend on $b(x_m, 0)$ only for $t > t_{mn}$.

\begin{figure*}[t!]
    \centering
    \includegraphics[width=\textwidth]{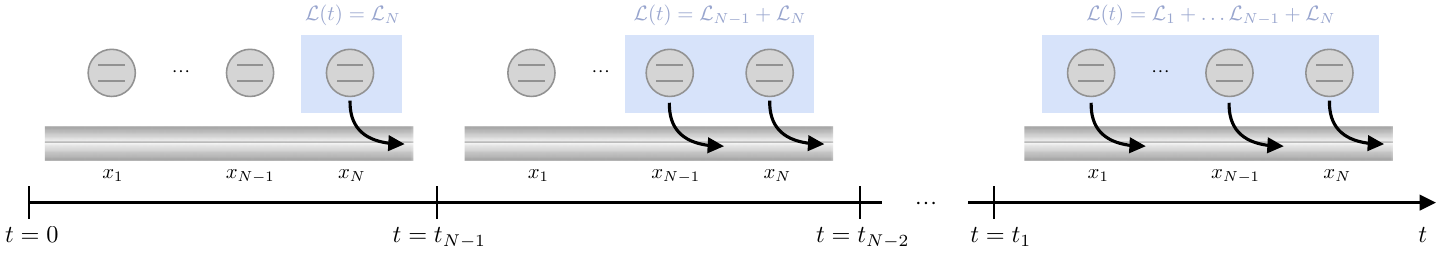}
    \caption{Schematic illustration of time-evolution in the time-shifted picture, viewed as the piecewise retardation-free dynamics of increasingly larger systems through the sequential inclusion of sites further upstream.}
    \label{fig:qme_illustration}
\end{figure*}

We can now derive a master equation for the reduced atomic density operator $\hat \rho(t) = \tr_\mathrm{wg}\{V(t)\chi(0)V^\dag(t)\}$, as follows.
Tracing Eq.~\eqref{eq:dotOtilde} with $\chi(0) = \rho(0)\otimes\ket{\vac}\!\bra{\vac}$, so that terms $\propto b(x, 0)$ and $\propto b^\dag(x, 0)$ vanish, and using the cyclic properties of the trace, we arrive at
\begin{equation}
    \partial_t \mean{\hat{O}(t)} = \Gamma \tr\left\{O \sum_n \left(\frac{1}{2}\Theta(t - t_n)[\sigma_n \hat\rho(t), \sigma^\dagger_n] + \sum_{m<n} \Theta(t - t_m) [\sigma_m \hat\rho(t), \sigma^\dagger_n] + \mathrm{\mathrm{H.c.}}\right)\right\} \,.
\end{equation}
Since this equation is valid for any atomic operator, $O$, it implies the master equation stated in the main text for the reduced atomic density operator, with a time-dependent Liouvillian $\mathcal{L}(t) = \sum_n \Theta(t - t_n)\mathcal{L}_n$, that involves an increasing number of emitters, as shown schematically in Fig.~\ref{fig:qme_illustration}.

When computing expectation values in the physical (un-shifted) frame, the time shift must be accounted for and reversed. For instance, one-time correlators between emitters take the form of two-time correlators in the time-shifted frame and vice versa. Specifically, we can define the correlator (as in the main text)
\begin{equation}
\label{eq:correlator_shifted}
    C(n,m,t)=\mean{\sigma_n^\dagger(t)\sigma_m(t-(n-m)\tau)}_0=\mathrm{tr}\{\sigma_n^\dagger\sigma_m\hat{\rho}(t+(N-n)\tau)\}\,.
\end{equation}
The occupations of the $n$-th emitter at time $t$ in the un-shifted frame are given by $C(n,n,t)$ (i.e. occupation of the $n$-th emitter at time $t+(N-n)\tau$ in the time-shifted frame). Furthermore, an expression for the site-$n$ emission rate $r_n(t)=-\partial_tC(n,n,t)$ can be computed by substituting $O=\sigma_n^\dagger\sigma_n$ in Eq.~\eqref{eq:dotOtilde} to obtain
\begin{equation}
    r_n(t) = \Gamma\left\{C(n,n,t)+2\sum_{m<n}\Theta(t-(n-m)\tau)\,\mathrm{Re}\,C(n,m,t)\right\}\,.
    \label{eq:emissionratecorr}
\end{equation}
From the above expressions, it is evident that in order to obtain the site-$N$ emission rate $r_N(t)$ up to some time $t$, we need to record one-time correlations in the time-shifted frame up to time $t+(N-1)\tau$.

Note that more general multi-time correlation functions in the time-shifted frame can also be computed.
For example, let us consider $\mean{\hat{A}(t')\hat{B}(t)}$, for $t' > t$, where $\hat{A}(t)$ and $\hat{B}(t)$ represent atomic operators in the time-shifted frame. 
It is easy to see that $\hat\rho_B(t) = \tr_\mathrm{wg}\{V(t)\hat{B}(t')\chi(0)V^\dagger(t)\}$ fulfills the same master equation as $\hat\rho(t)$, noting that $[b(x_N - vt, 0), \hat{B}(t')] = 0$ for all $n$.
Thus, $\mean{\hat{A}(t')\hat{B}(t)} = \tr\{A\mathcal{V}(t, t')[B\hat\rho(t')]\}$, for $t > t'$, with
\begin{equation}
    \mathcal{V}(t, t') = \mathcal{T}_\leftarrow\left\{\exp\left(\int_{t'}^t \mathcal{L}(s) \, ds\right)\right\} \,,
\end{equation}
where $\mathcal{T}_\leftarrow\left\{\dots\right\}$ denotes the usual time-ordered product of the operators inside the brackets.
Similarly, for $t' > t$, $\mean{\hat{A}(t')\hat{B}(t)} = \tr\{B\mathcal{V}(t', t)[\hat\rho(t)A]\}$.

\section{Details on numerical simulations}
In this section, we provide details on the numerical simulations employed in our work. We first review the exact simulations based on quantum trajectories with Matrix Product States (MPSs). This approach has previously proved successful in the context of WQED~\cite{manzoni_simulating_2017} owing to the fact that, as we will show, both the non-Hermitian Hamiltonian and the collective jump operator required for the trajectory simulations can be represented as Matrix Product Operators (MPOs) of fixed bond dimension. We then introduce the semi-classical Truncated Wigner Approximation (TWA) and present benchmarks establishing the accuracy of the TWA as compared to MPS simulations.

\subsection{Matrix Product States}
The action of the time-dependent Liouvillian can be re-written as $\mathcal{L}(t)\hat{\rho}(t)=i\left(\mathcal{H}(t)\hat{\rho}(t)-\hat{\rho}(t)\mathcal{H}^\dagger(t)\right)+S(t)\rho S^\dagger(t)$, where $S(t)=\sum_n\Theta(t-(N-n)\tau)\sigma_n$ is the permutation-invariant collective jump operator on the sites which are `active' at time $t$, and $\mathcal{H}$ is the non-Hermitian Hamiltonian~\cite{stannigel_driven-dissipative_2012,pichler_quantum_2015}
\begin{equation}
    \mathcal{H}(t)=-i\Gamma\sum_n\Theta(t-(N-n)\tau)\left\{\frac{1}{2}\sigma_n^\dagger\sigma_n-\sum_{m>n}\sigma_m^\dagger\sigma_n\right\}\,.
\end{equation}
Within the formalism of quantum trajectories~\cite{Carmichael1993,daley_quantum_2014,dalibard_wave-function_1992}, we divide the time span of interest into sufficiently small time intervals $\delta t$ and then assume a pure state $\ket{\psi(t)}$ of the atoms, which is evolved under the effective Hamiltonian $\mathcal{H}(t)$ as $\ket{\psi(t+\delta t)}\approx(1-i\mathcal{H}(t)\delta t)\ket{\psi(t)}$, punctuated by stochastic applications of the jump operator $S(t)$ (and subsequent normalisation of $\ket{\psi(t)}$). This gives rise to a single trajectory $\ket{\psi^{(j)}(t)}$. Observables (in the time-shifted picture) are then computed as stochastic averages over $N_t\gg 1$ trajectories obtained in this way, i.e.
\begin{equation}
    \mathrm{tr}\left(O\hat{\rho}(t)\right)\approx\frac{1}{N_t}\sum_{j=1}^{N_t}\bra{\psi^{(j)}(t)}O\ket{\psi^{(j)}(t)}\,.
\end{equation}
We can represent the state $\ket{\psi(t)}$ by an MPS and derive MPO representations for both the non-Hermitian Hamiltonian $\mathcal{H}(t)$ and the collective jump operator $S(t)$ with fixed bond dimension $D$ of the form
\begin{equation}
\label{eq:MPO}
    O(t)=\sum_{\alpha_1,\ldots,\alpha_{N-1}=1}^DI^{[1]}_{\alpha_0\alpha_1}\ldots I^{[n-1]}_{\alpha_{n-2}\alpha_{n-1}}\underbrace{O^{[n]}_{\alpha_{n-1}\alpha_n}\ldots O^{[N]}_{\alpha_{N-1}\alpha_N}}_{n\geq\lceil N-t/\tau\rceil}\qquad(\alpha_0=1,~\alpha_N=D)\,.
\end{equation}
Here, $O^{[n]}_{\alpha_{n-1}\alpha_n}$ is an operator-valued matrix acting on the $n$-th emitter, which for $\mathcal{H}$ and $S$ takes the form~\cite{mcculloch_density-matrix_2007,frowis_tensor_2010}
\begin{equation}
    \mathcal{H}^{[n]}=\begin{pmatrix}
        1 & -i\Gamma\sigma_n & -\dfrac{i\Gamma}{2}\sigma_n^\dagger\sigma_n \\ 0 & 1 & \sigma_n^\dagger \\ 0 & 0 & 1
    \end{pmatrix}\,,
    \qquad
    S^{[n]}=\begin{pmatrix}
        1 & \sigma_n \\ 0 & 1
    \end{pmatrix}\,,
\end{equation}
and $I^{[n]}_{\alpha_{n-1}\alpha_n}$ denotes the corresponding matrix with only identity operators on the diagonal. In fact, as noted in Ref.~\cite{manzoni_simulating_2017}, the first-order time-step operator $1-i\mathcal{H}(t)\delta t$ can also directly be represented by an MPO without increasing the bond dimension of the Hamiltonian MPO, through the substitution
\begin{equation}
    \mathcal{H}^{[1]} \to \begin{pmatrix}
        -i \delta t & -\delta t\Gamma\sigma_1 & 1-\dfrac{\delta t\Gamma}{2}\sigma_1^\dagger\sigma_1
    \end{pmatrix}\,.
\end{equation}

\begin{figure*}[t!]
    \centering
    \includegraphics[width=\linewidth]{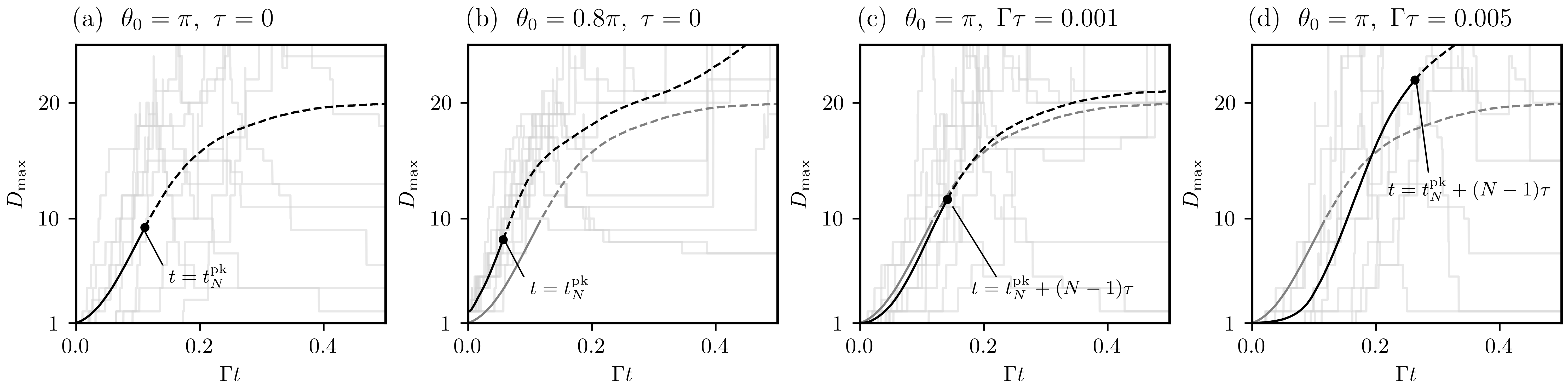}
    \caption{Evolution of the trajectory-averaged maximum bond dimension $D_\mathrm{max}$ for $N=20$ and (a) $\theta_0=\pi$ and $\tau=0$, (b) $\theta_0=0.8\pi$ and $\tau=0$, (c) $\theta_0=\pi$ and $\Gamma\tau=0.001$, (d) $\theta_0=\pi$ and $\Gamma\tau=0.005$. We indicate the peak emission time $t^\mathrm{pk}_n$ for (a),(b) and the time $t^\mathrm{pk}_n+(N-1)\tau$ for (c),(d). We underlay the evolution of the maximum bond dimension along individual trajectories in light gray in each subplot and in (b)-(d) we also plot the evolution of $D_\mathrm{max}$ from (a) in gray for reference.}
    \label{fig:mps_benchmarks}
\end{figure*}

The computational efficiency of the MPS simulations relates primarily to the growth of the maximum bond dimension of the MPS state representation, which we denote by $D_\mathrm{max}$. In Fig.~\ref{fig:mps_benchmarks}, we plot the time-evolution of $D_\mathrm{max}$ averaged over $N_t=20,000$ trajectories. For the fully-inverted initial state, we find that for $\tau=0$, $D_\mathrm{max}$ grows slowly and appears to plateau (see Fig.~\ref{fig:mps_benchmarks}a). Notably, $D_\mathrm{max}$ grows faster for partially-inverted states, however the peak emission also occurs earlier, so that ultimately the computational cost of computing dynamics up to time $t=t^\mathrm{pk}_N$ is not increased (see Fig.~\ref{fig:mps_benchmarks}b). For $\tau>0$, we find that the growth of $D_\mathrm{max}$ is initially slower than for $\tau=0$ but after some time grows faster, before also tapering off. While this effect is relatively insignificant for small $\tau$ (see Fig.~\ref{fig:mps_benchmarks}c), it becomes notable for larger $\tau$ (see Fig.~\ref{fig:mps_benchmarks}d). Therefore, for small $\tau$, the system sizes required to observe e.g. the plateau effect in the peak emission rate are much larger than can be reached using MPS (e.g. $N\approx 350$ for $\Gamma\tau=10^{-4}$ and $\theta_0=\pi$). On the other hand, for larger $\tau$, the requirements on $N$ are more attainable (e.g. $N\approx 30$ for $\Gamma\tau=10^{-2}$ and $\theta_0=\pi$), however in this case, the faster bond dimension growth becomes challenging. Additionally, the bond dimension growth on individual trajectories can be (much) faster than in the trajectory average (Fig.~\ref{fig:mps_benchmarks}).

\subsection{Truncated Wigner Approximation}
The TWA combines classical phase-space dynamics with Monte Carlo sampling of initial conditions to account for initial quantum fluctuations at leading order~\cite{polkovnikov_phase_2010,schachenmayer_many-body_2015,huber_phase-space_2021,huber_realistic_2022,mink_collective_2023,mink_hybrid_2022}. While the (discrete) TWA for spin systems was initially introduced for closed systems~\cite{schachenmayer_many-body_2015}, it has since been extended to account for local dissipation~\cite{huber_realistic_2022,huber_phase-space_2021} and, most recently, collective decay processes~\cite{mink_collective_2023,mink_hybrid_2022}. In our work, we employ in particular the method introduced in Ref.~\cite{mink_collective_2023}.

The equivalence of quantum dynamics on Hilbert space and phase space is formalised through the definition of a complete set of \emph{phase-point operators}~\cite{polkovnikov_phase_2010}. For a two-level system, one possible (continuous) parametrisation of the phase space is given by angular variables $\{\theta,\phi\}$, with phase-point operators chosen as $A(\theta,\phi)=(1+\vec{s}(\theta,\phi)\cdot\vec{\sigma})/2$, where $\vec{\sigma}=(\sigma^x,\sigma^y,\sigma^z)^T$ is the vector of Cartesian Pauli operators associated with the two-level system, and where
\begin{equation}
    \vec{s}(\theta,\phi)=\sqrt{3}\left(\sin\theta\cos\phi,\sin\theta\sin\phi,-\cos\theta\right)^T\,.
\end{equation}
An arbitrary operator $O$ acting on the Hilbert space of the two-level system can be mapped to its \emph{Weyl symbol} $W_O(\theta,\phi)$ (a function on phase space), and vice versa, through the relations
\begin{equation}
    W_O(\theta,\phi)=\mathrm{tr}\{A(\theta,\phi)O\}~\Leftrightarrow~O=\int d\Omega\,W_O(\theta,\phi)A(\theta,\phi)\,.
\end{equation}
Note that $\vec{s}(\theta,\phi)$ is the Weyl symbol of $\vec{\sigma}$. The above theory can be extended straightforwardly to a collection of $N$ two-level systems by defining the phase-point operators $A(\bm{\theta},\bm{\phi})=A_1(\theta_1,\phi_1)A_2(\theta_2,\phi_2)\ldots A_N(\theta_N,\phi_N)$.

The Weyl symbol associated with the (time-shifted) density operator, $W_{\hat{\rho}}(\bm{\theta},\bm{\phi})$, is the Wigner quasi-probability distribution. In principle, the full quantum dynamics of the density operator can be mapped onto a partial differential equation for $W_{\hat{\rho}}(\bm{\theta},\bm{\phi})$ under some correspondence rules, however this equation is in general exceedingly complicated, containing inifite-order derivatives with respect to the phase space variables. In Ref.~\cite{mink_collective_2023}, a set of alternative \emph{approximate} correspondence rules is introduced, which leads to a semiclassical equation of motion for the Wigner distribution. This equation is of the Fokker-Planck type and can therefore equivalently be captured by Stochastic Differential Equations (SDEs) for the phase space parameters $\{\bm{\theta},\bm{\phi}\}$~\cite{Gardiner2000}.

Applying these correspondence rules to the time-shifted picture master equation, we obtain $2N$ coupled SDEs of the form $d\theta_n=d\theta_n|_\mathrm{loc}+d\theta_n|_\mathrm{coh}+d\theta_n|_\mathrm{diss}$ and $d\phi_n=d\phi_n|_\mathrm{coh}+d\phi_n|_\mathrm{diss}$. The deterministic terms read
\begin{subequations}
\label{eq:twa}
\begin{align}
    &d\theta_n|_\mathrm{loc} = -\frac{\sqrt{3}\Gamma}{2}\,\Theta(t-(N-n)\tau)\left(\sin\theta_n-\frac{1}{\sqrt{3}}\cot\theta_n\right)dt \\
    &d\theta_n|_\mathrm{coh} = -\sqrt{3}\Gamma\sum_{m<n}\Theta(t-(N-m)\tau)\,\cos(\phi_n-\phi_m)\sin\theta_m\,dt \\
    &d\phi_n|_\mathrm{coh} = -\sqrt{3}\Gamma\sum_{m<n}\Theta(t-(N-m)\tau)\,\sin(\phi_n-\phi_m)\cot\theta_n\sin\theta_m\,dt
\end{align}
\end{subequations}
The stochastic terms are given by
\begin{subequations}
\begin{align}
    d\theta_n|_\mathrm{diss} &= \Gamma\,\Theta(t-(N-n)\tau)\,\left(\cos\phi_n\,dW_1-\sin\phi_n\,dW_2\right) \\
    d\phi_n|_\mathrm{diss} &= -\Gamma\,\Theta(t-(N-n)\tau)\,\cot\theta_n\left(\sin\phi_n\,dW_1+\cos\phi_n\,dW_2\right)
\end{align}
\end{subequations}
Note that for any $n$, the stochastic terms depend on the same two independent Wiener increments $dW_1$ and $dW_2$, reflecting the single-mode nature of the collective dissipation~\cite{mink_collective_2023}.

In practice, evolution of the phase space variables $\{\vec{\theta},\vec{\phi}\}$ under the stochastic equations gives rise to a single trajectory $\{\bm{\theta}^{(j)}(t),\bm{\phi}^{(j)}(t)\}$, subject to some initial conditions. For each such trajectory, we choose the initial conditions randomly, with probabilities dictated by the initial Wigner quasi-probability distribution. Observables (in the time-shifted picture) are then computed as stochastic averages over $N_t\gg 1$ trajectories obtained in this way, i.e.
\begin{equation}
    \mathrm{tr}\left(O\hat{\rho}(t)\right)=\overline{W_O(\bm{\theta},\bm{\phi})}\vert_t\approx\frac{1}{N_t}\sum_{j=1}^{N_t}W_O(\bm{\theta}^{(j)}(t),\bm{\phi}^{(j)}(t))
\end{equation}
Unless explicitly stated otherwise, throughout this work we choose $N_t=20,000$. Note that in the above expression, the evaluation of the Weyl symbol $W_O(\vec{\theta},\vec{\phi})$ may again involve applying the approximate correspondence rules. For instance, for the correlator $C(n,m,t)$ defined in Eq.~\eqref{eq:correlator_shifted}, we can straightforwardly obtain
\begin{subequations}
\begin{align}
    & C(n,n,t)= \frac{3}{4}\,\overline{\sin^2\theta_n}\vert_{t+(N-n)\tau}-\frac{\sqrt{3}}{2}\,\overline{\cos\theta_n}\vert_{t+(N-n)\tau} \\
    & C(n,m\neq n,t) = \dfrac{3}{4}\,\overline{e^{i(\phi_n-\phi_m)}\sin\theta_n\sin\theta_m}\vert_{t+(N-n)\tau}
\end{align}
\end{subequations}

\begin{figure*}[t!]
    \centering
    \includegraphics[width=\linewidth]{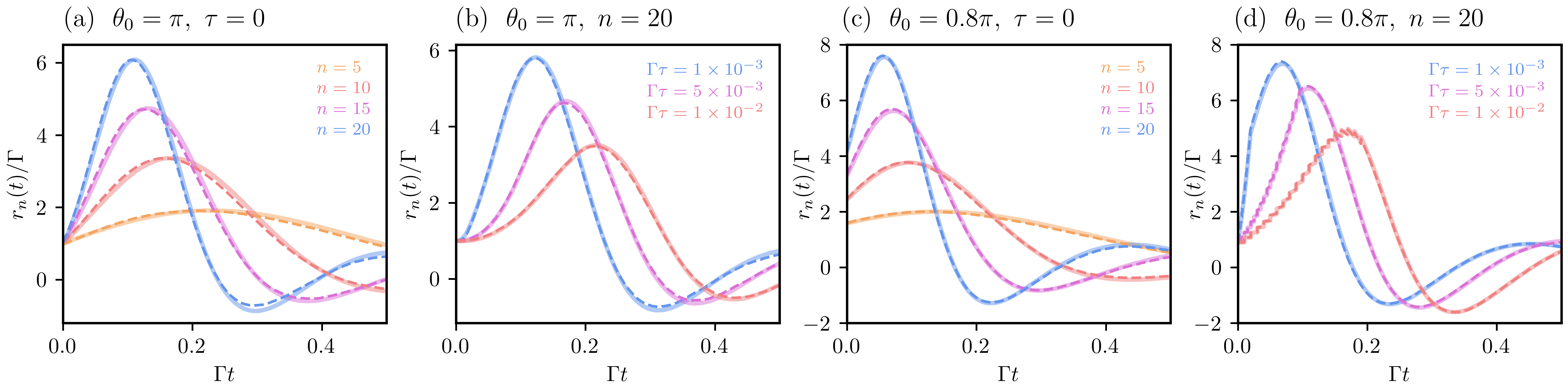}
    \caption{Time evolution of the site-$n$ emission rate $r_n(t)$, obtained using MPS (solid lines) and TWA (dashed lines). We fix the initial inversion angle $\theta_0$ and vary either the value of $n$ or retardation time $\tau$, as indicated above each panel.}
    \label{fig:twa_benchmarks}
\end{figure*}

Finally, we comment on the choice of initial conditions as sampled from the Wigner distribution for initial states considered in the main text. These are based on an equivalence between continuous and discrete parametrisations of the two-level phase space, as discussed in detail in Refs.~\cite{mink_hybrid_2022,mink_collective_2023}. Note that since these initial states are product states with each atom in the same state, the associated Wigner distributions take the general form $W_{\hat{\rho}}(\bm{\theta},\bm{\phi};t=0)=W_0(\theta_1,\phi_1)W_0(\theta_2,\phi_2)\ldots W_0(\theta_N,\phi_N)$. For the fully-inverted case $\theta_0=\pi$, we obtain the very simple form~\cite{mink_collective_2023}
\begin{equation}
    W_0(\theta,\phi)=\frac{1}{\sin\theta_\uparrow}\delta(\theta-\theta_\uparrow)\,,\quad\theta_\uparrow=\pi-\cos^{-1}\left(\frac{1}{\sqrt{3}}\right)\,,
\end{equation}
implying that the initial values of $\theta_n$ are fixed at $\theta_n=\theta_\uparrow$, while the initial values for each $\phi_n$ are sampled uniformly from $[0,2\pi)$. Away from the fully-inverted case, we similarly obtain the initial values $\theta_n=\theta(q_n)$ and $\phi_n=\phi(q_n,p_n)$ with $q_n,p_n$ ($n=1,\ldots,N$) each sampled uniformly from values $\{+1,-1\}$, and with
\begin{subequations}
\begin{align}
    &\theta(q) = \cos^{-1}\left(\frac{\cos\theta_0-q\sin\theta_0}{\sqrt{3}}\right) \\
    &\phi(q,p) = \tan^{-1}\big(p,q\cos\theta_0+\sin\theta_0\big)
\end{align}
\end{subequations}
In Fig.~\ref{fig:twa_benchmarks}, we compare the site-$n$ emission dynamics obtained using TWA and MPS for both a fully- and partially-inverted initial state and for various time delays $\tau$ and values of $n$. We find excellent agreement in all cases.

\section{Details on mean field theory}
In this section, we derive the continuum mean-field theory (MFT) given in the main text, and compare the mean-field dynamics with our TWA simulations for different parameters. We also derive an alternative discrete mean-field description starting from the master equation dynamics in the time-shifted picture, which we show to be equivalent to the continuum MFT in the retardation-free case and the continuum regime.

\subsection{Derivation of continuum MFT equation}
First, we derive the continuum mean-field equation, following broadly the derivations in Refs.~\cite{calajo_emergence_2022,mahmoodian_dynamics_2020}. We make the continuum approximation by introducing operators $\sigma(x)$ and $\sigma^z(x)$, with $\sigma_n=\sigma(x_n)d$ and $\sigma_n^z=\sigma^z(x_n)d$, which obey $[\sigma^\dagger(x),\sigma(y)]=2\sigma^z(x)\delta(x-y)$ and $[\sigma^z(x),\sigma(y)]=-\sigma(x)\delta(x-y)$. 
For $\Gamma\tau\ll1$, we approximate
\begin{equation}
    H \simeq -iv\int dx\,b^\dagger(x)\partial_xb(x)+\sqrt{\Gamma v}\int dx\,\left(\sigma^\dagger(x)b(x)+\mathrm{H.c.}\right)\,.
\end{equation}
for $t > 0$. Under a mean-field decoupling $\mean{AB}\approx\mean{A}\mean{B}$, we can then derive the equations of motion for $s(x,t)=\mean{\sigma(x)}_t$, $s^z(x,t)=\mean{\sigma^z(x)}_t$, and $\beta(x,t)=-i\mean{b(x)}_t$. Note that for initial states of the form considered in the main text, we can assume $s(x,t)$ to be real at all times, which implies for consistency that $\beta(x,t)$ will be real as well (see below). The MFT equations then read
\begin{subequations}
\begin{align}
    &\partial_t\beta(x,t) = -v\partial_x\beta(x,t)-\sqrt{\Gamma v}s(x,t) \label{eq:continuous_mft}\\
    &\partial_ts(x,t) = -2\sqrt{\Gamma v}s^z(x,t)\beta(x,t) \\
    &\partial_ts^z(x,t) = 2\sqrt{\Gamma v}s(x,t)\beta(x,t)
\end{align}
\end{subequations}
The last two of these equations admit the solution $s(x,t)=\frac{1}{2d}\sin\theta(x,t)$ and $s^z(x,t)=-\frac{1}{2d}\cos\theta(x,t)$, where the inversion angle function $\theta(x,t)$ is related to $\beta(x,t)$ according to
\begin{equation}
\label{eq:continuous_mft_relation}
    \theta(x,t)=2\sqrt{\Gamma v}\int_0^tdt'\,\beta(x,t')\quad\implies\quad\beta(x,t)=\frac{1}{2\sqrt{\Gamma v}}\partial_t\theta(x,t)\,.
\end{equation}
Substituting the solutions for $s(x,t)$ and $\beta(x,t)$ into Eq.~\eqref{eq:continuous_mft}, we obtain a partial differential equation for $\theta(x,t)$. To make the arguments of $\theta(x,t)$ dimensionless, we re-scale $x\to x/d$ and $t\to \Gamma t$. We then obtain the final equation for $\theta(x,t)$ quoted in the main text by dividing across by $\Gamma v/d$ in the resulting equation. Quenching on the coupling at time $t=0$ implies the initial condition $\theta(x,t=0)=\theta_0$, where $\theta_0$ is the initial inversion angle. We obtain the second boundary condition $\theta(x=0,t)=\theta_0$ by integrating with respect to $x$ and taking $x\to0$. This second condition essentially reflects the fact that the continuum MFT only accounts for the collective decay, therefore the atom furthest upstream cannot decay.

\begin{figure*}[t!]
    \centering
    \includegraphics[width=\linewidth]{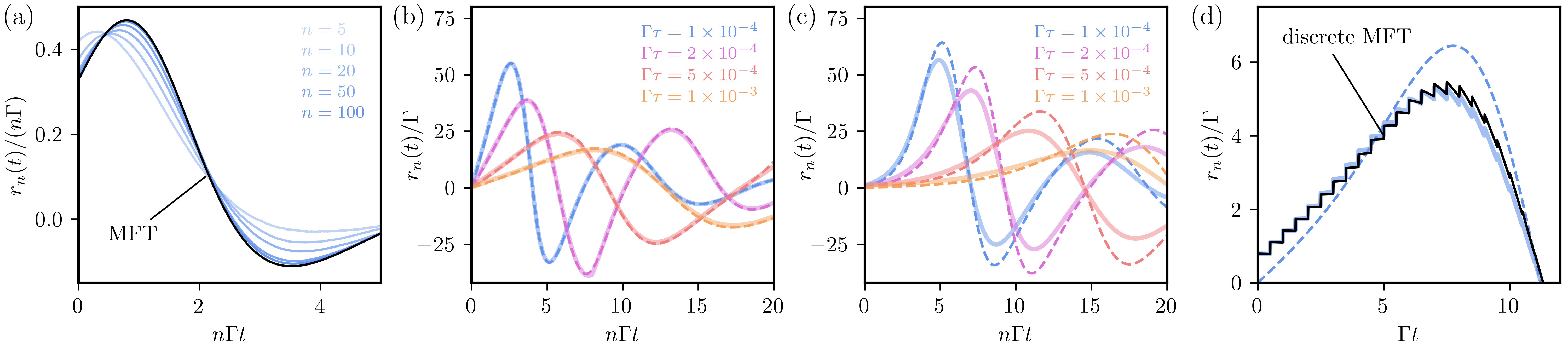}
    \caption{(a) Site-$n$ emission rates for $\theta_0=0.7\pi$ and increasing $n$, computed using the TWA (blue lines), compared with the continuum MFT solution to the Sine-Gordon equation (black line). (b),(c) Comparison of local emission dynamics for $n=200$, computed using TWA (solid lines) and continuum MFT (dashed lines), for various time delays $\tau$ and (b) $\theta_0=0.6\pi$ and (c) $\theta_0=0.9\pi$. (d) Comparison of local emission rate $r_n(t)$ for $n=50$, $\theta_0=0.7\pi$, and $\Gamma\tau=10^{-2}$, computed using TWA (blue solid line), continuum MFT (blue dashed line), and discrete MFT (black line).}
    \label{fig:mft_benchmarks}
\end{figure*}

In Fig.~\ref{fig:mft_benchmarks}, we compare the MFT against TWA simulations. As noted in the main text, for $\tau=0$, the MFT equation has the symmetric sollution $\theta(x,t)=u(xt)$, which implies the asymptotic scaling $r_n(t)=(n/m)r_m(nt/m)$. In Fig.~\ref{fig:mft_benchmarks}a, we show that this scaling is indeed approached by the TWA dynamics in the limit of large $n,m$. Note that agreement with the TWA is, in some sense, unsurprising; the initial quantum fluctuations, which distinguish the TWA from the MFT, scale as $\sim1/\sqrt{n}$~\cite{ma_quantum_2011}, so that in the limit of very large $n$ and $\tau=0$, the TWA and MFT dynamics should be expected to be essentially equivalent. For $\tau>0$, we find excellent agreement with the TWA for smaller initial inversion angles $\theta_0$ (see Fig.~\ref{fig:mft_benchmarks}b). For larger $\theta_0$, however, the agreement can be seen to be much worse (see Fig.~\ref{fig:mft_benchmarks}c), reflecting the increased importance of the initial quantum fluctuations to the dynamics, which are not captured by the MFT~\cite{Agarwal1974}.

As noted in the main text, for $\tau>0$ we can also obtain an asymptotic mean-field equation as $x\to\infty$ by dropping the spatial derivative from the MFT equation. The solution $\theta_\infty(t)$ to this equation can be obtained analytically as $\theta_\infty(t)=\mathrm{am}(\varphi_0-t/k\sqrt{\Gamma\tau}|k^2)$, where $\varphi_0=\mathrm{F}(\theta_0/2|k^2)$ and $k=1/\sin(\theta_0/2)$. Here, $\mathrm{F}(\cdot|\cdot)$ and $\mathrm{am}(\cdot|\cdot)$ denote the incomplete elliptic integral of the first kind and the Jacobi amplitude, respectively (analytically continued to $k^2>1$). The solution is oscillatory with a period $T=2k\sqrt{\Gamma\tau}K(1-k^2)$, where $K(\cdot)$ is the elliptic integral of the first kind.

\subsection{Discrete MFT in time-shifted picture}
The most striking example of a regime where the continuum MFT breaks down is shown in Fig.~\ref{fig:mft_benchmarks}d; for large $\tau$ and relatively large $\theta_0$, the continuum MFT not only fails to accurately approximate the peak emission but also cannot capture the granularity of the atom array which manifests in a noticeable step-wise increase of the emission rate. Moreover, the TWA shows the expected initial independent decay $r_N(t=0)=\Gamma\sin^2(\theta_0/2)$, while the continuum MFT, as noted above, does not account for on-site decay. For the sake of completeness, we note that to better approximate this regime within a mean-field approach, we can also derive a MFT starting directly from the master equation in the time-shifted frame without making a continuum approximation.

Starting from the time-dependent master equation, the equations of motion for the (time-shifted) spin expectation values $s_n(t)=\mean{\hat{\sigma}_n(t)}$ and $s_n^z(t)=\mean{\hat{\sigma}_n^z(t)}$ can be derived under a mean-field decoupling as
\begin{subequations}
\label{eq:discrete_MFT}
\begin{align}
    \partial_ts_n(t) &= -\frac{\Gamma}{2}\Theta(t-(N-n)\tau)\,s_n(t) + 2\Gamma\sum_{m<n}\Theta(t-(N-m)\tau)\,s_n^z(t)s_m(t)  \label{eq:mft_spin} \\ 
    \partial_ts_n^z(t) &= -\Gamma\,\Theta(t-(N-n)\tau)\left(s_n^z(t)+\frac{1}{2}\right)-2\Gamma\sum_{m<n}\Theta(t-(N-m)\tau)\,\mathrm{Re}\big(s_n^*(t)s_m(t)\big)
\end{align}
\end{subequations}
As we see in Fig.~\ref{fig:mft_benchmarks}d, this discrete MFT shows arguably better agreement with the TWA simulations in the regime where the continuum approximation is not particularly valid. Note that Eqs.~\eqref{eq:discrete_MFT} can also be shown to be equivalent to the TWA equations, with the caveats that (i) the stochastic terms must be dropped, since they vanish in the statistical expectation values ($\overline{dW}=0$), (ii) the on-site terms in the TWA equations differ, since the mean-field decoupling is not applied to them, and (iii) the TWA equations are re-scaled by a factor of $\sqrt{3}$ arising from the conventions in the definition of the phase-point operators.

While the discrete MFT at the level of Eqs.~\eqref{eq:discrete_MFT} bears little resemblance to the continuum MFT derived in the previous section, in the retardation-free case we can actually establish a direct connection between the two. In the limit of large $n$, i.e. for sites sufficiently far downstream a sufficiently long chain of atoms, the contributions from on-site terms in Eqs.~\eqref{eq:discrete_MFT}, which induce dynamics on timescales $\sim\Gamma^{-1}$, will be negligible compared to the collective terms, associated with dynamics on timescales $\sim(n\Gamma)^{-1}$. Once these terms are dropped, it is easy to see that $(s_n^z)^2+\abs{s_n}^2$ represents a constant of the motion. Using this to substitute for $s_n^z$ in Eq.~\eqref{eq:mft_spin} allows us to decouple the equation of motion for $s_n$ from that for $s_n^z$ at each $n$, whereby
\begin{equation}
    \partial_ts_n(t)=-\Gamma\sqrt{1-4s_n^2(t)}\sum_{m<n}s_m(t)\,.
\end{equation}
Transforming to angular variables as $s_n(t)=\frac{1}{2}\sin\theta_n(t)$, we find that $\partial_t\theta_n(t)=-\Gamma\sum_{m<n}\sin\theta_m(t)$. We now make the continuum approximation by defining the bivariate function $\theta(x,t)$ with dimensionless arguments, from which $\theta_n(t)=\theta(n,\Gamma t)$. Thus, we obtain
\begin{equation}
    \partial_t\theta(x,t)\simeq-\int_0^x dy\,\sin\theta(y,t)\quad\implies\quad\partial_x\partial_t\theta(x,t)+\sin\theta(x,t)=0\,.
\end{equation}

\section{Fitting correlations at peak emission}
\begin{figure*}[h!]
    \centering
    \includegraphics[width=\linewidth]{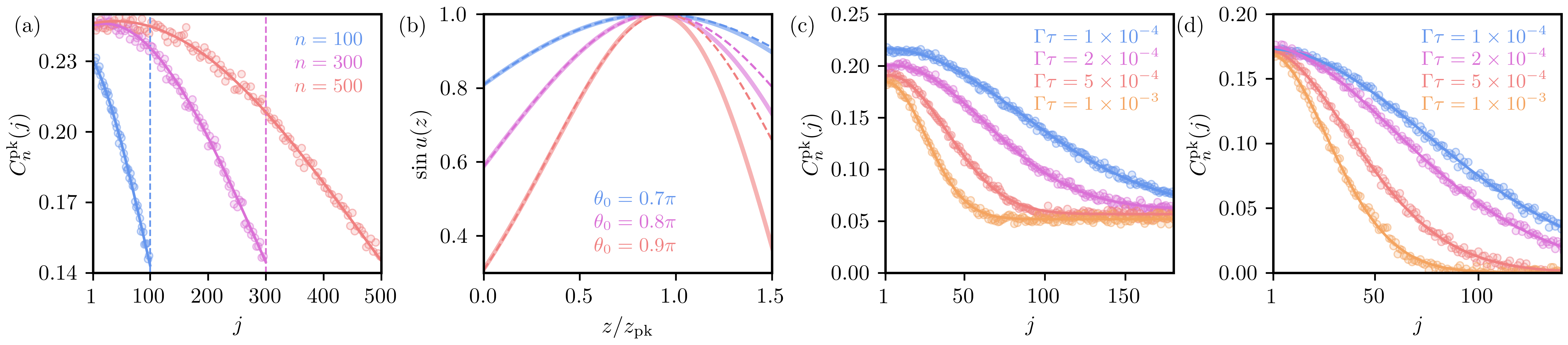}
    \caption{(a) Correlator $C^\mathrm{pk}_n(j)$ for a partially-inverted initial state with $\theta_0=0.8\pi$ in the retardation-free limit. We plot the correlations computed using TWA (markers), as well as Gaussian fits (solid lines), for various $n$. (b) Evolution of $\sin u(z)$, where $u(z)$ is the solution to the retardation-free mean-field equation (solid lines) and Gaussian fits (dashed lines) for various initial inversion angles $\theta_0$. (c) Correlations at peak emission $C^\mathrm{pk}_n(j)$ for a partially-inverted initial state with $\theta_0=0.9\pi$ and $n=200$, for different $\tau$. We plot the correlations computed using TWA (markers), as well as Gaussian fits (solid lines). (d) Correlations at peak emission $C^\mathrm{pk}_n(j)$ for a fully-inverted initial state and $n=150$, for different $\tau$. We plot the correlations computed using TWA (markers), as well as KWW fits (solid lines).}
    \label{fig:correlations}
\end{figure*}

Here, we discuss the empirical fitting of the correlator at peak emission, $C^\mathrm{pk}_n(j)\equiv C(n,n-j,t^\mathrm{pk}_n)$. We first consider the retardation-free case, for which we find that $C^\mathrm{pk}_n(j)$ is well-captured by a Gaussian fit function (see Fig.~\ref{fig:correlations}a). Some qualitative understanding on the accuracy of a Gaussian fit can be obtained by noting that, in the mean-field picture, the correlator is given by $C^\mathrm{pk}_n(j)\approx C_0\sin u((1-j/n)z_\mathrm{pk})$ with $C_0=\frac{1}{4}\sin u(z_\mathrm{pk})$, where we have defined $z_\mathrm{pk}$ as the value of $z$ which maximises the mean-field emission rate in the retardation-free case,
\begin{equation}
    r(z)=\frac{\sin(z)}{2z}\int_0^z dz'\,\sin u(z')\,,
\end{equation}
and where $u(z)$ denotes the solution to the Sine-Gordon equation (see main text). As we show in Fig.~\ref{fig:correlations}b, for $z\leq z_\mathrm{pk}$, $\sin u(z)$ is well-described by a Gaussian fit, i.e. $\sin u(z)\approx e^{-(z-c_1)^2/c_2^2}$. This implies $C^\mathrm{pk}_n(j)\approx C_0 e^{-(jd-c_3)^2/c_4^2}$ with $c_3=nd(1-c_1/z_\mathrm{pk})$ and $c_4=c_2nd/z_\mathrm{pk}$. Note that $c_4$ corresponds with $\xi_n$ defined in the main text, and $c_4\sim n d$ consistent with the scaling of $\xi_n$ which we observe numerically.

For $\tau>0$, we find that for partially-inverted states, a Gaussian fit is still a good approximation, with an additional constant offset to account for the uniform correlations within the ``light cone'' described in the main text (see Fig.~\ref{fig:correlations}c). For the fully-inverted initial state, on the other hand, we find that a better fit is given by the more general compressed exponential function $C_n^\mathrm{pk}(j)=C_0e^{-(jd/\xi_n(t))^\alpha}$ introduced in the main text (see Fig.~\ref{fig:correlations}d). Note that in the literature, this function is also known as the \emph{Kohlrausch–Williams–Watts} (KWW) function~\cite{kohlrausch_theorie_1854,williams_non-symmetrical_1970}, and has found applications in a wide range of relaxation processes as an empirical fit function for various observables~\cite{adanlete_adjanoh_compressed_2011,gregorin_dynamics_2022,palmer_models_1984,talantsev_relaxation_2018}, including multi-time correlators like the ones we study here~\cite{madsen_beyond_2010,ruta_atomic-scale_2012}.

\bibliography{references_papers.bib, references_books.bib}